\PassOptionsToPackage{table}{xcolor}
\documentclass[acmtog,screen,nonacm]{acmart}
\AtBeginDocument{%
  }

\definecolor{rankone}{RGB}{255,153,153}
\definecolor{ranktwo}{RGB}{255,204,153}
\definecolor{rankthree}{RGB}{255,248,173}
\newcommand{\rankonecell}[1]{\cellcolor{rankone!80}\textbf{#1}}
\newcommand{\ranktwocell}[1]{\cellcolor{ranktwo!80}#1}
\newcommand{\rankthreecell}[1]{\cellcolor{rankthree!80}#1}
\DeclareRobustCommand{\rankonelegend}[1]{\setlength{\fboxsep}{1pt}\colorbox{rankone!80}{\strut\textbf{#1}}}
\DeclareRobustCommand{\ranktwolegend}[1]{\setlength{\fboxsep}{1pt}\colorbox{ranktwo!80}{\strut#1}}
\DeclareRobustCommand{\rankthreelegend}[1]{\setlength{\fboxsep}{1pt}\colorbox{rankthree!100}{\strut#1}}

\newif\ifshowmidfig
\showmidfigtrue
\setcopyright{none}
\acmJournal{TOG}
\acmSubmissionID{2616}

\begin{document}

\title{OctaOctree Neural Radiosity for Real-time Glossy Material Rendering}

%\author{Anouymous Author}

\author{Jierui Ren}
\email{jerry@stu.pku.edu.cn}
\orcid{0009-0000-3556-4402}
\affiliation{%
  \institution{Peking University}
  \city{Beijing}
  \country{China}
}

\author{Haojie Jin}
\email{jhj@pku.edu.cn}
\affiliation{%
  \institution{Peking University}
  \city{Beijing}
  \country{China}
}

\author{Bo Pang}
\email{bo98@stu.pku.edu.cn}
\affiliation{%
  \institution{Peking University}
  \city{Beijing}
  \country{China}
}

\author{Meng Gai}
\email{gaimeng@pku.edu.cn}
\affiliation{%
  \institution{Peking University}
  \city{Beijing}
  \country{China}
}

\author{Fei Zhu}
\email{feizhu@pku.edu.cn}
\affiliation{%
  \institution{Peking University}
  \city{Beijing}
  \country{China}
}

\author{Yisong Chen}
\email{chenyisong@pku.edu.cn}
\affiliation{%
  \institution{Peking University}
  \city{Beijing}
  \country{China}
}

\author{Sheng Li}
\email{lisheng@pku.edu.cn}
\authornote{Corresponding author.}
\affiliation{%
  \institution{Peking University}
  \city{Beijing}
  \country{China}
}


% \renewcommand{\shortauthors}{Anouymous Author}

\begin{abstract}
Modeling high-frequency outgoing radiance distributions remains a fundamental challenge in global illumination, especially for glossy and specular materials. Existing neural-based radiance caching methods commonly rely on positional feature encodings or spatially organized caches, which makes it difficult to represent sharp directional radiance variations without increasing the model complexity or sampling cost.
To address this challenge, we propose OctaOctree, an efficient spatial-angular radiance representation for global illumination. OctaOctree organizes outgoing radiance with an adaptive octree in 3D space, and associates each spatial node with an octahedral directional map. By coupling the spatial hierarchy with direction-dependent storage, our representation allocates fine spatial resolution to local illumination and visibility changes, while using coarser spatial levels with richer angular resolution to capture glossy and specular radiance distributions.
This design embeds a reflectance-aware spatial-angular prior directly into the radiance representation, reducing the burden on neural networks or reconstruction modules to recover high-frequency view-dependent effects from positional features alone. As a result, OctaOctree provides a compact and expressive neural encoding for a wide range of indirect illumination effects, from diffuse interreflection to sharp glossy reflections. Experiments demonstrate that our method produces high-quality, direction-aware global illumination with single network query at primary intersections, achieving improved fidelity and real-time performance compared with baseline neural radiosity and radiance caching approaches.
\end{abstract}

\begin{CCSXML}
<ccs2012>
<concept>
<concept_id>10010147.10010371.10010372.10010374</concept_id>
<concept_desc>Computing methodologies~Ray tracing</concept_desc>
<concept_significance>500</concept_significance>
</concept>
</ccs2012>
\end{CCSXML}

\ccsdesc[500]{Computing methodologies~Ray tracing; Neural networks}

\keywords{Neural scene representation, global illumination, neural radiosity, glossy materials}
\begin{teaserfigure}
  \includegraphics[width=\textwidth]{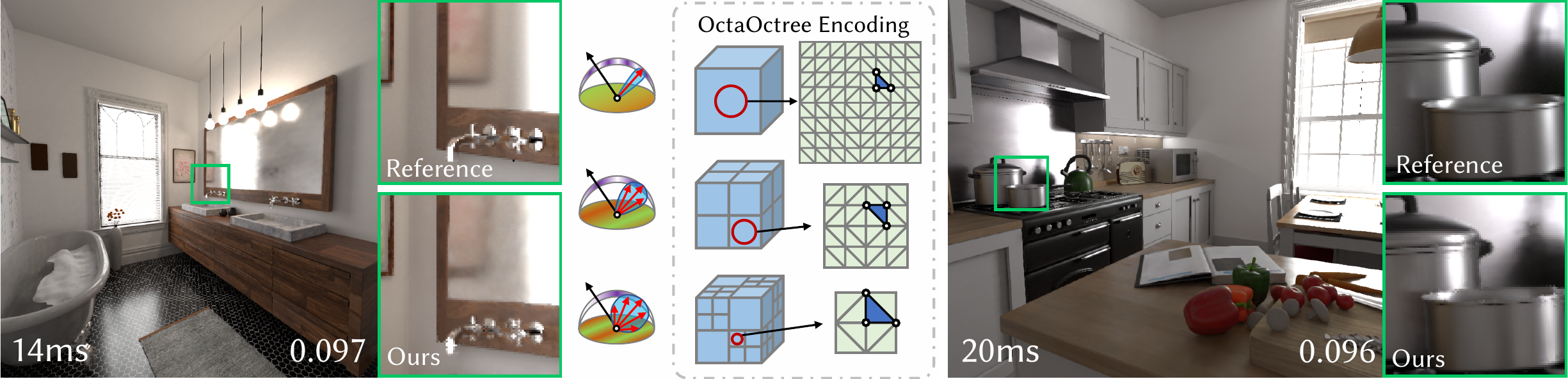}
  \caption{Two scenes rendered with our OctaOctree neural radiosity. Our method efficiently captures high-frequency view-dependent effects for glossy BSDF lobes of different roughness with complementary spatial-angular feature budget. The images are rendered at $1024\times1024$ and $1280\times720$ resolution with Mitsuba3 renderer. We show the comparisons in the cropped area between reference images rendered with path tracing, and the results of our method.}
  \label{fig:teaser}
\end{teaserfigure}


\maketitle

\section{Introduction}

Accurately reproducing indirect light remains one of the central costs of physically based rendering. In offline rendering, Monte Carlo integration and path tracing provide a general and physically faithful way to evaluate global illumination across complex geometry, materials, and lighting. Yet this accuracy comes at the price of repeatedly tracing and integrating large numbers of light transport paths. For real-time and interactive applications, where similar transport queries must be answered again and again across frames and viewpoints, this repeated computation quickly becomes the bottleneck.

Neural global illumination methods aim to reduce that cost by replacing part of the transport computation with learned prediction. One line of work stays in image space and learns to clean up sparse Monte Carlo renderings through denoising and feature-aware filtering~\cite{chaitanya2017interactive,bako2017kpcn}. A second line predicts indirect light more directly from local shading cues such as G-buffers, visibility surrogates, or low-cost direct-light features~\cite{nalbach2017deepshading,thomas2017deepillumination,xin2022lightweight}. A third line moves the representation into scene space and learns reusable radiance functions or caches that can be queried during rendering~\cite{muller2021nrc,hadadan2021neural}. These approaches differ in where the learned prior is applied, but all share the same motivation: to amortize the cost of indirect illumination without giving up too much fidelity.

Among these directions, scene-space methods are especially attractive because they reuse the transport solution itself. Neural Radiance Caching and Neural Radiosity~\cite{muller2021nrc,hadadan2021neural} learn a reusable radiance representation for a fixed scene, which enables efficient queries at render time and typically produces smoother indirect illumination than repeatedly estimating all secondary transport from scratch.

Their main limitation is directional representation. Existing NR/NRC-style methods usually combine strong positional encoding with only compact directional inputs, which is often adequate for diffuse transport but not for glossy outgoing radiance. As a result, the cache itself neither model sharp angular variation explicitly nor exploit the spatial distribution patterns that glossy radiance exhibits across nearby surface points, so reflections tend to become noisy or overly blurred.

Neural Cone Radiosity (NCR)~\cite{ren2025ncr} takes an important step toward this problem by introducing cone-based encoding for glossy transport. However, it still relies on multiple ray intersections to evaluate each glossy query, which increases computation substantially, and the resulting estimates may still retain a small amount of residual noise.

To address this limitation, we propose \textbf{OctaOctree}, an efficient spatial-angular radiance representation for scene-specific neural radiosity. The core idea is to encode outgoing radiance with an adaptive octree in 3D space and to associate each spatial node with an octahedral directional map. By coupling spatial hierarchy with directional storage, OctaOctree directly embeds a reflectance-aware spatial-angular prior into the representation. Fine spatial levels capture local illumination and visibility variation, while coarser levels preserve richer angular resolution for glossy and specular transport.

% Unlike previous approaches that rely mainly on positional feature encodings or point-wise neural evaluation, OctaOctree exposes directional radiance structure explicitly at the representation level. This substantially reduces the burden on the neural parameterization to infer high-frequency view-dependent effects from position alone. As a result, OctaOctree serves as a compact and expressive neural encoding for a wide range of indirect illumination phenomena, from diffuse interreflection to sharp glossy reflections.

Unlike previous approaches that rely mainly on positional encodings or point-wise neural evaluation, OctaOctree exposes directional radiance structure at the representation level, reducing the burden on the network to infer high-frequency view-dependent effects implicitly.

We implement OctaOctree in a scene-specific neural radiosity setting and evaluate it on scenes with diffuse, glossy, and specular materials. Experiments show that the proposed representation improves the fidelity of view-dependent global illumination with single network query at primary intersections, while achieving better performance compared with baseline neural radiosity and radiance-caching-style approaches.

Overall, our main technical contributions are:
\begin{itemize}
\item We propose \textbf{OctaOctree}, a spatial-angular radiance representation that combines an adaptive octree in 3D space with octahedral directional maps.
\item We introduce a complementary allocation of spatial and angular resolution that is better suited to glossy and specular transport than purely positional neural representations.
\item We demonstrate that this reflectance-aware cache representation improves the quality and efficiency of scene-specific neural radiosity and radiance caching for direction-dependent global illumination.
\end{itemize}

\section{Related Work}

\subsection{Neural Rendering for Global Illumination}

Neural methods for global illumination span several levels of representation. A large body of work operates in image space, where convolutional networks or learned filters reconstruct clean global illumination from noisy Monte Carlo renderings and auxiliary buffers~\cite{chaitanya2017interactive,bako2017kpcn,vogels2018denoising,icsik2021interactive}. Related screen-space approaches learn shading or indirect-lighting effects directly from G-buffers, direct lighting, or other rasterization features~\cite{nalbach2017deepshading,thomas2017deepillumination,granskog2020compositional,diolatzis2022active}. Recent extensions further combine super-resolution and frame extrapolation to improve efficiency in real-time pipelines~\cite{wu2023extrass,zhong2023fusesr}. These methods are highly practical, but because they act after projection into screen space, they do not explicitly store a reusable scene-space transport solution.

Another line of work predicts indirect illumination more directly from scene or shading features. Examples include lightweight learned GI predictors for interactive rendering and methods designed for dynamic lighting conditions~\cite{xin2022lightweight,gao2022ngi}. These approaches amortize transport effectively, but they largely ask the network to regress the final radiance signal from compact features. As a result, highly view-dependent or multi-bounce effects can still be difficult to preserve when the underlying representation is too compressed.

% Most closely related to our work are scene-space transport methods. Neural Importance Sampling and Neural Control Variates demonstrated that neural networks can model transport-relevant distributions for Monte Carlo rendering~\cite{muller2019neural,muller2020neural}, while Neural Radiance Caching (NRC)~\cite{muller2021nrc} and Neural Radiosity~\cite{hadadan2021neural} showed that scene-specific neural representations can learn reusable radiance solutions directly in 3D space. Recent work has also applied similar ideas to path guiding~\cite{dong2023neural,dong2024efficient}. Our method follows this scene-space direction, but shifts the focus from network prediction alone to the cache representation itself: OctaOctree explicitly organizes outgoing radiance jointly over position and direction, which is particularly beneficial for glossy transport.

Most closely related to our work are scene-space transport methods. Neural Importance Sampling and Neural Control Variates demonstrated that neural networks can model transport-relevant distributions for Monte Carlo rendering~\cite{muller2019neural,muller2020neural}, while Neural Radiance Caching (NRC)~\cite{muller2021nrc} and Neural Radiosity~\cite{hadadan2021neural} showed that scene-specific neural representations can learn reusable radiance solutions directly in 3D space. Subsequent work further extended neural radiosity to dynamic scenes~\cite{su2024dynamic} and more compact scene representations~\cite{su2025vertex}. Recent work has also applied similar ideas to path guiding~\cite{dong2023neural,dong2024efficient}. However, existing NRC/NR-RHS-style methods usually query learned radiance at secondary intersections, so glossy results still inherit stochastic noise from sampled transport. Meanwhile, NR-style primary-hit prediction relies on compact directional inputs, forcing the MLP to infer high-frequency directional variation implicitly. Our method follows the same scene-space formulation, but explicitly organizes outgoing radiance jointly over space and direction, allowing glossy directional structure to be stored inside the cache and queried once at the primary intersection.

\subsection{Structured Neural Scene Encodings}

Coordinate-based neural scene representations have become a standard formulation for modeling spatially varying signals in graphics. NeRF popularized the view that scene appearance can be represented as a continuous function of position and direction~\cite{mildenhall2021nerf}, and Fourier or positional encodings improved the ability of multilayer perceptrons to fit high-frequency functions~\cite{tancik2020fourier}. This paradigm has since been extended far beyond novel-view synthesis to radiance prediction, geometry optimization, and dynamic scene modeling.

To improve efficiency, many works introduced explicit spatial structure. Neural Sparse Voxel Fields and KiloNeRF reduce the burden on a single large MLP by distributing scene content across sparse or local subnetworks~\cite{liu2020nsvf,reiser2021kilonerf}. Other methods replace heavy neural backbones with directly optimized spatial features, as in PlenOctrees, Plenoxels, Direct Voxel Grid Optimization, TensoRF, and K-Planes~\cite{yu2021plenoctrees,fridovich2022plenoxels,sun2022direct,chen2022tensorf,tang2022compressible,fridovich2023k}. Instant-NGP further showed that multiresolution hash encodings can greatly accelerate scene-specific optimization~\cite{muller2022instantngp}, and related work extended structured encodings to dynamic lighting or deformable transport fields~\cite{coomans2024real,su2024dynamic,zheng2024neural}.

Several works also improved directional fidelity and anti-aliasing in neural representations. Mip-NeRF, Tri-MipRF, and Zip-NeRF account for finite spatial footprints instead of purely infinitesimal samples~\cite{barron2021mip,hu2023tri,barron2023zip}. Ref-NeRF and NeRFReN enrich the representation for reflective appearance by introducing more structured modeling of normals, roughness, or reflection components~\cite{verbin2022ref,guo2022nerfren}.

On the representation side, Octahedron Environment Maps introduced octahedral parameterization as an efficient way to encode directional data~\cite{engelhardt2008octahedron}, and DDGI later adopted octahedral maps to store probe irradiance and visibility for real-time global illumination~\cite{majercik2019ddgi}. Our work is motivated by a similar observation, but in the context of neural radiosity: instead of storing diffuse irradiance, we use similar directional layout to store learned outgoing-radiance features, and couple it with spatial adaptivity for glossy transport.

\subsection{Glossy Indirect Illumination}

Glossy indirect illumination remains challenging because near-specular transport is strongly view-dependent and often contains narrow, high-frequency directional lobes. Earlier real-time solutions typically relied on precomputation or filtering. Precomputed radiance transfer targeted low-frequency transport and distant lighting~\cite{sloan2002precomputed}, while later methods considered more general BRDFs and interreflections~\cite{xu2014practical}. In interactive settings, low-sample path tracing combined with spatiotemporal reconstruction remains a common strategy~\cite{schied2017spatiotemporal}, but glossy reflections are still prone to instability, blur, or residual noise.

Probe- and cache-based methods improve temporal stability by storing reusable lighting information in scene space. Glossy Probe Reprojection and Efficient Light Probes show that carefully designed caches can support glossy transport more robustly than purely image-space methods~\cite{rodriguez2020glossyprobe,guo2022efficient}. Within neural GI, NRC and NIRC query learned scene-space predictors at secondary ray hits or along transport paths~\cite{muller2021nrc,dereviannykh2025nirc}. This avoids building an explicit glossy module, but the final result still inherits the variance of stochastic secondary-ray evaluation. Neural Radiosity moves closer to direct evaluation at primary shading points~\cite{hadadan2021neural}, yet compact directional inputs make very sharp glossy effects difficult to fit faithfully. Neural Cone Radiosity further improves NR for glossy materials by introducing cone-based directional aggregation~\cite{ren2025ncr}, but it still relies on tracing multiple rays and performing repeated ray-surface intersections to evaluate glossy transport.

Recent neural GI systems have begun to address richer reflection structure more explicitly. NeLT and LightFormer incorporate stronger conditioning on scene layout or light-source structure to better reproduce view-dependent illumination~\cite{zheng2023nelt,ren2024lightformer}, while Dual-Band Feature Fusion separates and fuses different frequency bands for multi-frequency reflections~\cite{mo2025dualfusion}. These methods show the importance of representation design for glossy transport. Our approach pursues the same goal from a cache-construction perspective: instead of adding a specialized reflection branch, we store outgoing radiance in an adaptive octree with octahedral directional maps, yielding a scene-specific neural radiosity cache that is better aligned with the spatial-angular structure of glossy global illumination.

\section{Method}

\subsection{Overview}

\begin{figure}[t]
\centering
\includegraphics[width=\linewidth]{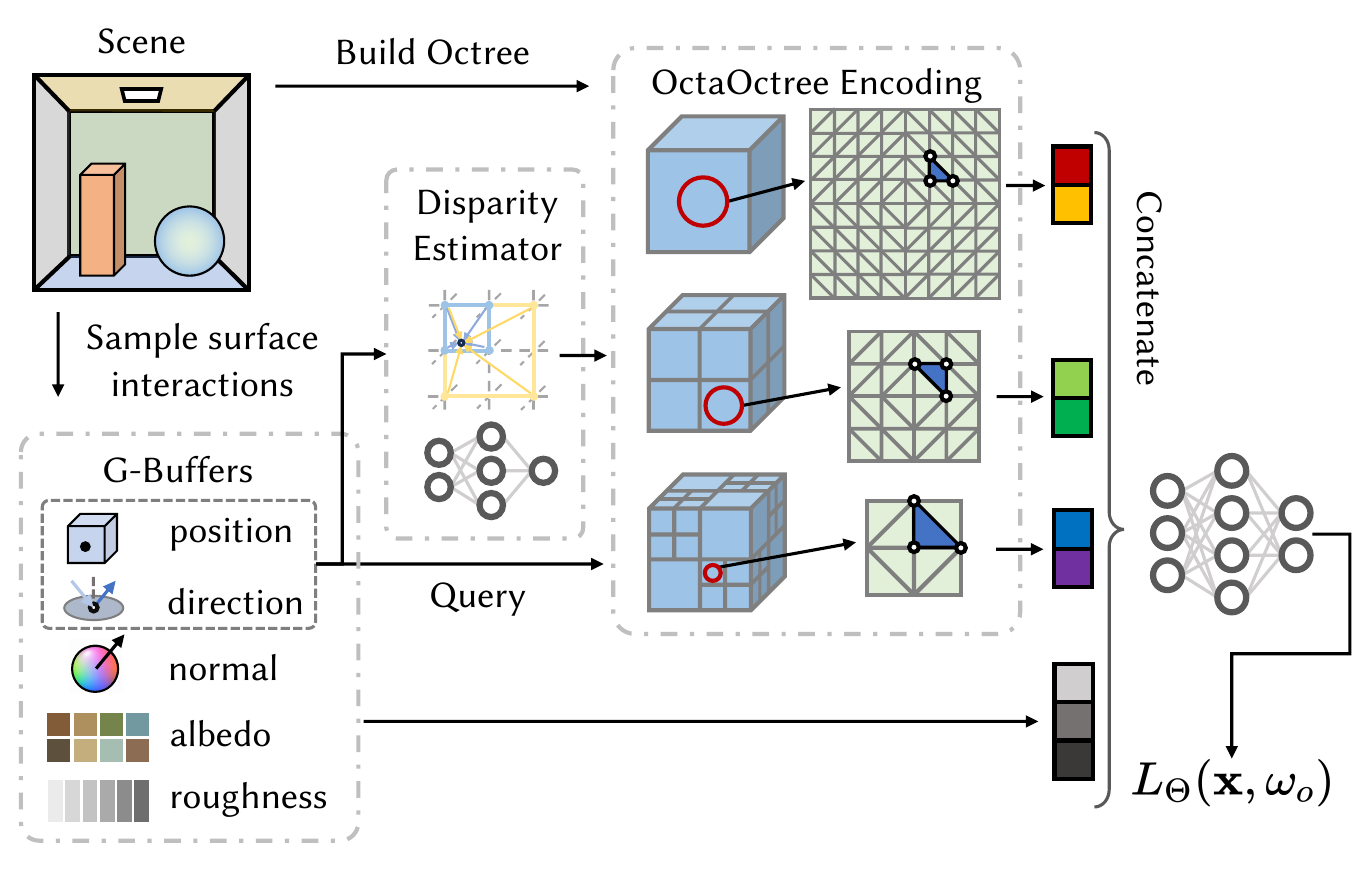}
\caption{
The overall pipeline of our method. We represent outgoing radiance with an OctaOctree spatial-angular cache, where each octree cell stores an octahedral directional map. A compact predictor estimates a disparity value for each shading point, which is used to shift the octahedral lookup direction before interpolation. The network takes concatenated features from all levels of the OctaOctree and G-Buffers as input, and predicts the final outgoing radiance.
}
\label{fig:model}
\end{figure}

We consider a scene with geometry, materials, and lighting, and aim to learn a scene-specific outgoing radiance field for interactive rendering. Our method follows the neural radiosity formulation of learning a function that predicts outgoing radiance at surface points, but replaces the usual point-wise spatial encoding with an explicit spatial-angular cache. The overall pipeline is illustrated in Figure~\ref{fig:model}.

The key idea is to represent outgoing radiance jointly over position and direction. We organize the scene with a multi-resolution octree in 3D space, and associate each spatial cell with an octahedral directional map. We refer to this representation as \emph{OctaOctree}. In addition, we train a compact Instant-NGP-style predictor that estimates a disparity value for a given surface point and reflection direction, which is used to shift the octahedral lookup direction before interpolation. During training, the OctaOctree features, the disparity predictor, and a lightweight decoder are optimized together by enforcing the rendering equation. During rendering, a visible shading point queries the learned OctaOctree and the decoder predicts the final outgoing radiance in a single forward pass.

\subsection{Rendering Equation and Neural Radiosity}

The outgoing radiance at a surface point $\mathbf{x}$ along direction $\omega_o$ is governed by the rendering equation
\begin{equation}
L_o(\mathbf{x},\omega_o)
=
L_e(\mathbf{x},\omega_o)
+
\int_{\Omega}
f_r(\mathbf{x},\omega_i,\omega_o)
L_i(\mathbf{x},\omega_i)
\left|n_{\mathbf{x}} \cdot \omega_i\right|
\mathrm{d}\omega_i,
\end{equation}
where $L_e$ is emitted radiance, $f_r$ is the BSDF, $n_{\mathbf{x}}$ is the surface normal, and $L_i$ denotes incident radiance from direction $\omega_i$.

For opaque surfaces, the incident radiance is the outgoing radiance of the next visible surface point:
\begin{equation}
L_i(\mathbf{x},\omega_i)
=
L_o(\mathbf{x}'(\omega_i),-\omega_i),
\end{equation}
where $\mathbf{x}'(\omega_i)$ is the first intersection reached by tracing a ray from $\mathbf{x}$ along $\omega_i$. Neural Radiosity~\cite{hadadan2021neural} exploits this self-consistency by treating the left-hand side (LHS), $L_o(\mathbf{x},\omega_o)$, as the prediction to be learned, and the right-hand side (RHS) of the rendering equation as its supervisory target.

Let $\hat{L}_\Theta(\mathbf{x},\omega_o)$ denote the radiance predicted by a learnable scene representation with parameters $\Theta$. Neural Radiosity discretizes the RHS with Monte Carlo samples and uses it to supervise the LHS:
\begin{equation}
\tilde{L}(\mathbf{x},\omega_o)
=
L_e(\mathbf{x},\omega_o)
+
\frac{1}{M}
\sum_{m=1}^{M}
\frac{
f_r(\mathbf{x},\omega_i^m,\omega_o)
\hat{L}_\Theta(\mathbf{x}'^m,-\omega_i^m)
\left|n_{\mathbf{x}} \cdot \omega_i^m\right|
}{
p(\omega_i^m)
},
\end{equation}
where $\omega_i^m \sim p(\omega)$ are sampled incident directions and $\mathbf{x}'^m$ are the corresponding secondary intersections. The network cache is then updated by minimizing the discrepancy between $\hat{L}_\Theta(\mathbf{x},\omega_o)$ and $\tilde{L}(\mathbf{x},\omega_o)$. In this sense, Neural Radiosity is self-supervised: the network is trained by using the discretized RHS to supervise the LHS, rather than by regressing to precomputed ground-truth radiance values.

This formulation is attractive because it learns a reusable transport solution for a scene. However, existing neural radiosity and radiance caching methods usually rely on point-wise spatial encodings together with compact directional inputs~\cite{hadadan2021neural,muller2021nrc}. This is often sufficient for diffuse transport, but becomes limiting for glossy materials, where outgoing radiance varies sharply with direction. In such cases, the network must infer narrow directional lobes from features that are primarily organized in space, which often leads to oversmoothed reflections or unstable secondary-ray estimates. Neural Cone Radiosity alleviates this issue by introducing cone-based glossy encoding~\cite{ren2025ncr}, but it still requires multiple ray-surface intersections per glossy query and may retain a small amount of residual noise. Our method addresses this limitation at the representation level by explicitly storing direction-dependent radiance structure in OctaOctree.

\subsection{Octahedral Directional Interpolation}

\begin{figure}[t]
\centering
\includegraphics[width=\linewidth]{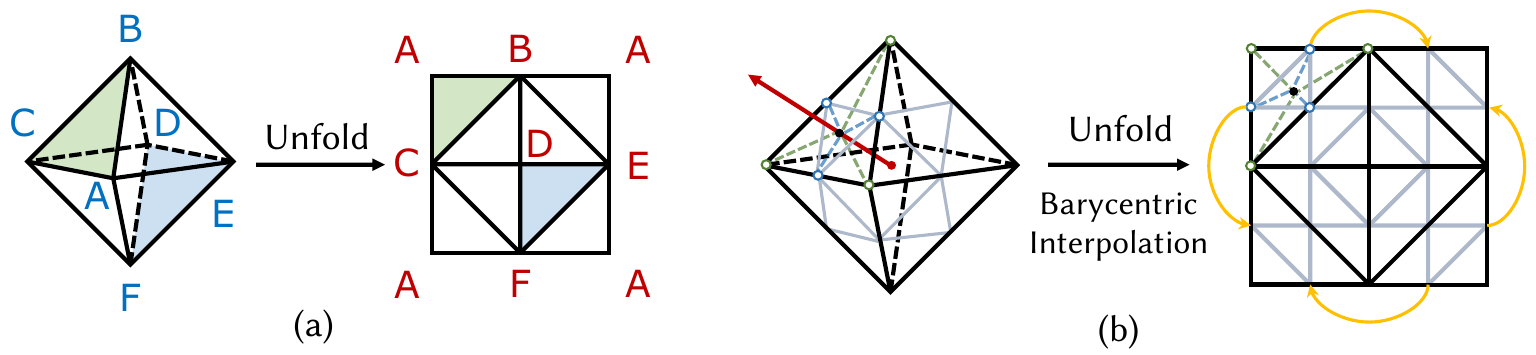}
\caption{
(a) An octahedron and corresponding unfolded map. (b) For a queried direction, we locate the containing triangle on the unfolded grid and compute barycentric weights for its three vertices. Interpolation is conducted on multiple levels of the octahedral maps with subdivided triangles. The edge vertices are shared along adjacent faces, so the interpolation is continuous across the spherical domain.
}
\label{fig:octaUnfold}
\end{figure}

To represent directional variation efficiently, each spatial cell stores radiance features on an octahedral map (See Figure~\ref{fig:octaUnfold}(a)). This directional parameterization follows Octahedron Environment Maps~\cite{engelhardt2008octahedron}, and is related to its later use in DDGI for storing probe irradiance and visibility~\cite{majercik2019ddgi}. In our setting, however, we use a similar layout to store learned outgoing-radiance features rather than diffuse lighting quantities. Given a unit direction $\omega=(\omega_x,\omega_y,\omega_z)$, we first project it onto the unfolded octahedral domain:
\begin{equation}
\mathbf{u}
=
\frac{(\omega_x,\omega_y)}
{\left|\omega_x\right|+\left|\omega_y\right|+\left|\omega_z\right|}.
\end{equation}
If $\omega_z < 0$, the lower hemisphere is folded onto the upper one by
\begin{equation}
\mathbf{u}
\leftarrow
\left(
\left(1-\left|u_y\right|\right)\mathrm{sign}(u_x),
\left(1-\left|u_x\right|\right)\mathrm{sign}(u_y)
\right).
\end{equation}
The resulting coordinates are remapped to $[0,1]^2$ and used to access the multi-resolution directional grid. At any fixed directional resolution, each face of the octahedron is uniformly subdivided into smaller equilateral triangles. After unfolding, these cells become a regular tessellation of right isosceles triangles on the 2D octahedral map. Therefore, the directional lookup at a given resolution reduces to locating one such triangle and interpolating the three features stored at its vertices.

For a queried direction, we locate the containing triangle on the unfolded grid and compute barycentric weights for its three vertices (See Figure~\ref{fig:octaUnfold}(b)). Let $u_1,u_2,u_3$ be the triangle vertices and $b_1,b_2,b_3$ the barycentric weights. The directional feature inside one spatial cell is then
\begin{equation}
\mathbf{h}^{\mathrm{dir}}(\omega)
=
\sum_{k=1}^{3}
b_k(\omega)\mathbf{t}_{u_k},
\end{equation}
where $\mathbf{t}_{u_k}$ is the learned feature stored at directional vertex $u_k$.

\begin{figure}[t]
\centering
\includegraphics[width=\linewidth]{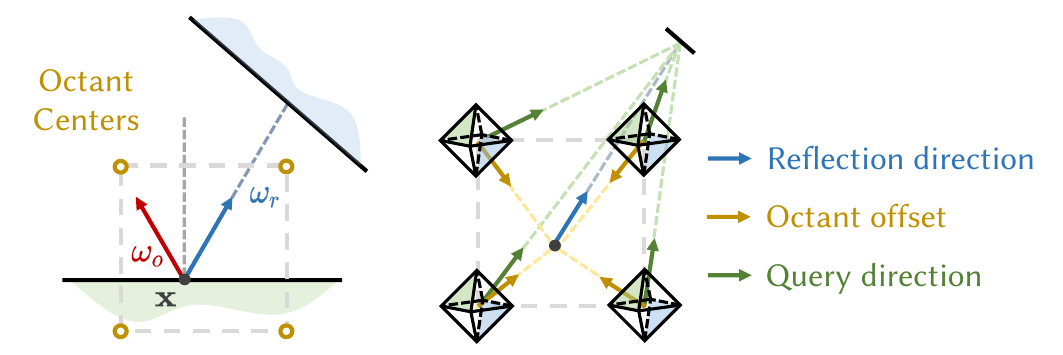}
\caption{
The direction shift mechanism in our method. The predicted disparity shifts the lookup direction before interpolation, aligning it with the reflected scene content.
}
\label{fig:dirShift}
\end{figure}

Using the raw outgoing direction $\omega_o$ for lookup is often suboptimal on glossy surfaces, because the dominant reflected energy is typically centered around the mirror reflection direction rather than the viewing direction itself. We therefore define the mirror reflection direction $\omega_r$ at the shading point and predict a disparity value
\begin{equation}
\hat{d}(\mathbf{x},\omega_r)=g_{\psi}(\mathbf{x},\omega_r),
\end{equation}
where $g_{\psi}$ is a compact multi-resolution hash-grid network in the style of Instant-NGP~\cite{muller2022instantngp}. The predicted disparity determines a virtual reflected point
\begin{equation}
\mathbf{y}(\mathbf{x},\omega_r)=\mathbf{x}+\frac{\omega_r}{\hat{d}(\mathbf{x},\omega_r)}.
\end{equation}
For each neighboring octree cell centered at $c_j$, we use the shifted direction
\begin{equation}
\tilde{\omega}_j=
\frac{\mathbf{y}(\mathbf{x},\omega_r)-c_j}{\|\mathbf{y}(\mathbf{x},\omega_r)-c_j\|_2}=\frac{(\mathbf{x}-c_j)\hat{d}(\mathbf{x},\omega_r) + \omega_r}{\|(\mathbf{x}-c_j)\hat{d}(\mathbf{x},\omega_r) + \omega_r\|_2},
\end{equation}
for octahedral lookup instead of directly using $\omega_o$ or $\omega_r$ (See Figure~\ref{fig:dirShift}).

This octahedral interpolation with direction shift (or reprojection) is a key component of our method. It enables smooth directional queries while still allowing the representation to preserve high-frequency angular structure, which is especially important for glossy transport. Unlike ray-traced depth, which changes discontinuously at reflection visibility boundaries, the learned disparity is optimized only through the radiance reconstruction objective. It therefore tends to represent a local smooth reprojection field that improves cache interpolation rather than a geometrically accurate hit distance (See Figure~\ref{fig:disparity_visualization}).

Compared with Neural Cone Radiosity~\cite{ren2025ncr}, which improves glossy transport through cone-based aggregation over multiple traced intersections, our method keeps the inference procedure much simpler. The spatial-angular structure is absorbed into the learned OctaOctree itself, so at render time we do not need additional multi-ray traversal or clustering to evaluate glossy reflection.

\subsection{OctaOctree Encoding}

OctaOctree combines a spatial octree with directional octahedral maps. Let $l \in \{1,\dots,K\}$ denote a hierarchy level. Each level partitions scene space into sparse octree cells, and every non-empty cell stores a learnable octahedral feature map. The spatial resolution increases from coarse to fine, while the directional resolution changes in the opposite direction to reflect a complementary spatial-angular reuse pattern. A dense 3D directional cache can be viewed as a special case of spatial-angular storage. Our contribution is not merely storing directional radiance, but coupling angular resolution with adaptive spatial hierarchy and learned reprojection.

This design is motivated by the rendering equation: outgoing radiance can be viewed as incident radiance convolved with the BSDF. As surface roughness decreases, the BSDF support becomes narrower, so the valid reuse neighborhood in direction shrinks and the outgoing signal becomes increasingly high-frequency in angle. At the same time, the reflected signal tends to vary more smoothly across nearby surface points under the direction shift, so the valid reuse neighborhood in space expands. Rougher transport exhibits the opposite behavior, with broader angular averaging and more localized spatial dependence due to geometry, visibility, and illumination variation. OctaOctree follows this trade-off by combining larger spatial cells with finer directional discretization, and finer spatial cells with coarser directional discretization.

Each OctaOctree entry stores a $C$-dimensional trainable feature. For a query point $\mathbf{x}$, we first locate non-empty octants from the eight neighboring octree cells at level $l$ and compute standard trilinear weights $w_j^{\mathrm{sp}}(\mathbf{x})$. Inside each neighboring cell, the shifted direction $\tilde{\omega}_j$ is evaluated on the octahedral map using the barycentric interpolation described above. The level-$l$ feature is
\begin{equation}
\mathbf{h}_l(\mathbf{x},\omega_o)
=
\sum_{j=1}^{8}
\frac{\mathbf{1}(c_j)w_j^{\mathrm{sp}}(\mathbf{x})}{\sum_{p=1}^{8}\mathbf{1}(c_j)w_p^{\mathrm{sp}}(\mathbf{x})}
\sum_{k=1}^{3}
w_{j,k}^{\mathrm{dir}}(\tilde{\omega}_j)
\mathbf{T}_{l,c_j,u_{j,k}},
\end{equation}
where $c_j$ is one of the neighboring spatial cells, $u_{j,k}$ are the three directional vertices selected in that cell, $\mathbf{1}(\cdot)$ indicates the octree cell is non-empty, and $\mathbf{T}_{l,c_j,u_{j,k}} \in \mathbb{R}^{C}$ is the corresponding trainable feature.

The full OctaOctree query concatenates features from all levels:
\begin{equation}
\mathbf{h}(\mathbf{x},\omega_o)
=
\left[
\mathbf{h}_1(\mathbf{x},\omega_o),
\mathbf{h}_2(\mathbf{x},\omega_o),
\ldots,
\mathbf{h}_K(\mathbf{x},\omega_o)
\right].
\end{equation}
The resulting queried representation therefore has dimension $KC$.

Compared with a purely positional encoding, this representation makes the directional structure of outgoing radiance explicit. The network no longer has to infer all glossy behavior from position and a low-dimensional direction code alone; instead, it receives features that are already organized according to the spatial-angular structure of outgoing radiance distribution.

\subsection{Network Structure}

The network built on top of OctaOctree is intentionally lightweight. It contains two small neural components in addition to the OctaOctree parameters: the disparity predictor $g_{\psi}$ and a radiance decoder $f_{\phi}$. The predictor $g_{\psi}$ uses a multi-resolution hash encoding followed by a tiny MLP to regress $\hat d(\mathbf{x},\omega_r)$. The OctaOctree itself serves as the main scene representation, while the decoder is implemented as an MLP that maps queried features to outgoing radiance. Due to the explicit spatial-angular structure in the OctaOctree, the decoder can be much more compact compared to vanilla NR~\cite{hadadan2021neural} and still capture high-frequency view-dependent effects.

Given a shading point $(\mathbf{x},\omega_o)$, we concatenate the OctaOctree query with local surface attributes:
\begin{equation}
\mathbf{g}(\mathbf{x},\omega_o)
=
\left[
\mathbf{x},
\omega_r,
n,
a,
\alpha
\right],
\end{equation}
where $\mathbf{x}$ is the shading position, $\omega_r$ is the mirror reflection direction, $n$ is the surface normal, $a$ is the diffuse albedo, and $\alpha$ is the roughness. The final prediction is
\begin{equation}
\hat{L}_\Theta(\mathbf{x},\omega_o)
=
f_\phi\left(\mathbf{h}(\mathbf{x},\omega_o),\mathbf{g}(\mathbf{x},\omega_o)\right),
\end{equation}
where $\Theta$ includes the OctaOctree features together with the parameters of both neural modules.

This design follows the intuition of neural radiosity and radiance caching: the expensive part of transport is absorbed into a scene-specific representation, while the neural networks remain small and fast. By making angular variation explicit inside each octree cell, and by using the predicted disparity to align directional lookup with reflected scene content, OctaOctree reduces the amount of directional reasoning that must be synthesized implicitly by the decoder.

\subsection{Optimization and Rendering}

We optimize the scene-specific representation by enforcing the rendering equation at randomly sampled surface interactions. For each training sample $(\mathbf{x},\omega_o)$, the network predicts $\hat{L}_\Theta(\mathbf{x},\omega_o)$, and a Monte Carlo estimate of the right-hand side provides the supervisory target. Following standard neural radiosity practice, we detach the RHS target and the normalization term from backpropagation, which stabilizes optimization.

Our training loss is a relative $L_2$ objective:
\begin{equation}
\mathcal{L}
=
\left\|
\frac{
\hat{L}_\Theta(\mathbf{x},\omega_o)
-
\mathrm{sg}\!\left[\tilde{L}(\mathbf{x},\omega_o)\right]
}{
\mathrm{sg}\!\left[
\hat{L}_\Theta(\mathbf{x},\omega_o)
+
\tilde{L}(\mathbf{x},\omega_o)
+
\epsilon
\right]
}
\right\|_2^2,
\end{equation}
where $\epsilon$ is a small constant. The optimization is performed independently for each scene.

At inference time, rendering reduces to standard primary visibility plus one OctaOctree query and one decoder evaluation per visible shading point. No additional denoising or reconstruction stage is required. In this sense, our method can be viewed as a scene-specific neural radiosity cache whose main contribution is the OctaOctree spatial-angular encoding.

\section{Results}

\subsection{Implementation Details}

We evaluate OctaOctree on path-traced scenes rendered with Mitsuba~3. To support real-time inference, we implement OctaOctree lookup and interpolation with a custom CUDA kernel, while the neural modules are executed with the tiny-cuda-nn backend used in Instant-NGP~\cite{muller2022instantngp}. Our octree implementation is inspired by the structure of o-cnn~\cite{wang2017cnn}. All experiments are conducted on an desktop with an NVIDIA RTX 3090 GPU and Intel Core i7-10700K.

Unless otherwise stated, all reported images are rendered in linear HDR radiance and tone mapped only for visualization. We report mean absolute percentage error (MAPE), LPIPS~\cite{zhang2018lpips} along with inference time. MAPE is measured in linear radiance space, and LPIPS is evaluated on tone-mapped images. The results are evaluated against path-traced references generated with 102400 samples per pixel. The inference time covers OctaOctree and network query, as well as ray-traced G-buffer extraction from Mitsuba.

Our implementation uses $K=8$ OctaOctree levels. The spatial resolution ranges from $(2^2)^3$ to $(2^9)^3$, while the octahedral directional resolution changes in the opposite direction. Since directional features are stored at map vertices, the corresponding vertex resolutions range from $2^8+1$ at the coarsest level to $2^1+1$ at the finest level. Each OctaOctree entry stores a 4-dimensional feature, producing a 32-dimensional feature vector after concatenating all levels. In addition to the OctaOctree itself, we use a compact Instant-NGP-style network with 2 hidden layers of 64 neurons to predict disparity from position and mirror reflection direction $\omega_r$, and a radiance decoder implemented as a 4-hidden-layer MLP with width 128. We train each scene for 20{,}000 steps. Each batch initially contains 65{,}536 left-hand-side samples, and every LHS sample is paired with $M=32$ Monte Carlo samples for the right-hand-side estimate. Following Dynamic Neural Radiosity~\cite{su2024dynamic}, after every quarter of the training schedule we halve the number of LHS samples and double $M$, progressively shifting computation from spatial coverage toward a lower-variance RHS target. The whole training process takes roughly 1 hour to fully converge, depending on scene complexity.

\subsection{Main Comparison}

We compare against four representative baselines that reflect different ways of handling glossy global illumination.

\paragraph{Neural Cone Radiosity (NCR)}
We compare against Neural Cone Radiosity~\cite{ren2025ncr}, which explicitly models glossy transport through cone-based directional aggregation. This is the most closely related glossy-oriented baseline.

\paragraph{Neural Radiosity (NR)}
This baseline follows the direct primary-hit formulation of Neural Radiosity~\cite{hadadan2021neural}. It predicts outgoing radiance at visible surface points from a scene-specific neural representation, but does not explicitly encode directional information.

\paragraph{OIDN}
As an image-space reference, we compare against low-sample (4 spp) path tracing followed by Intel Open Image Denoise. This baseline tests whether a strong post-process denoiser can recover the same reflected detail without an explicit scene-space radiance cache.

\paragraph{Deferred NR/NRC}
This baseline represents methods that query the learned cache only at the second bounce, in the spirit of the RHS formulation of Neural Radiosity and Neural Radiance Caching~\cite{hadadan2021neural,muller2021nrc}. It preserves more of the stochastic transport process, but in glossy regions the final result still depends on noisy secondary-ray evaluation.

The scenes are chosen to cover transport effects that are difficult to reconstruct from sparse sampling, including diffuse interreflection, glossy reflections, high-frequency reflected details, and sharp highlights. For each scene, geometry, materials, and lighting remain fixed, and we optimize a dedicated scene-specific radiance field.

Table~\ref{tab:main_quality} reports quantitative quality across all test scenes, and Figure~\ref{fig:main_comparison} shows representative visual comparisons. Our method improves the reconstruction of glossy reflections and reflected details while achieving significant and consistent performance gain on all scenes.

The differences are most visible on glossy surfaces. OIDN improves perceptual smoothness but cannot explicitly recover scene-space angular structure. It also shows apparent flickering artifacts during interactive rendering. Deferred NR/NRC avoids strong deterministic bias, yet its dependence on sampled secondary rays leaves visible noise in difficult view-dependent regions. NR provides stable direct predictions at the primary hit, but compact directional inputs are often insufficient for narrow reflected lobes. NCR handles glossy transport more explicitly, yet it still relies on ray-traced prefiltering which introduces minor noise and performance overhead. In contrast, OctaOctree stores outgoing radiance jointly over space and direction, allowing the network to operate on features that are already aligned with the structure of glossy transport. Please refer to the supplementary video for more visual comparisons and interactive renderings.

\ifshowmidfig
\begin{figure*}[t]
\centering
\includegraphics[width=\textwidth,height=0.90\textheight,keepaspectratio]{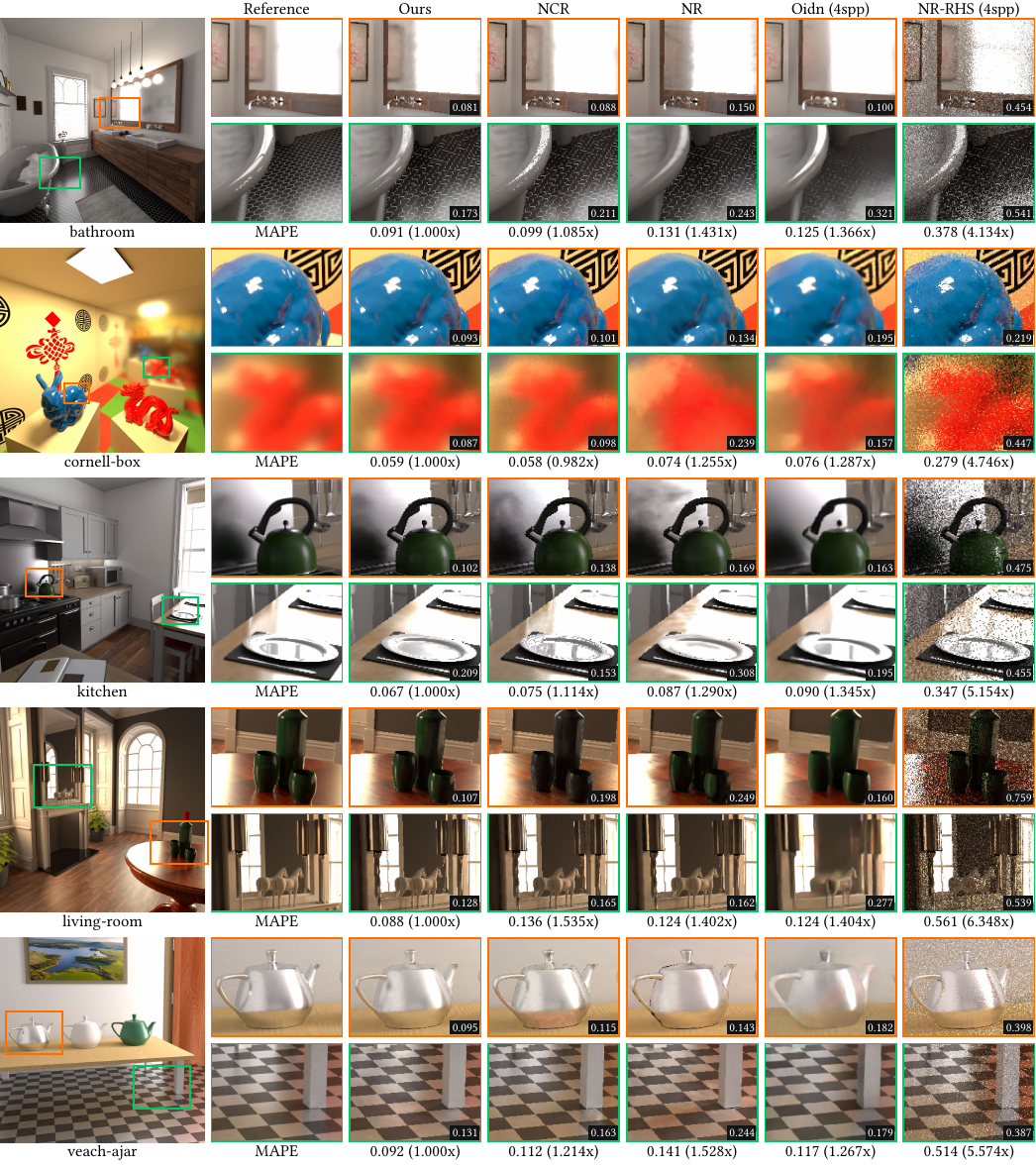}
\caption{
Visual and qualitative comparisons on the main test scenes. We present the rendering results of our method (\emph{Ours}), Neural Cone Radiosity~\cite{ren2025ncr} (\emph{NCR}), vanilla Neural Radiosity(\emph{NR}), OIDN(\emph{OIDN}), and deferred radiance cache queried for RHS bounces (\emph{NR-RHS}). Main visual comparison. OIDN removes much of the Monte Carlo noise but tends to smooth reflected structure in screen space. NR-RHS retains the stochastic nature of secondary-bounce evaluation and may leave residual noise. NR is cleaner, but often oversmooths narrow glossy lobes. NCR improves glossy transport, while our method produces more accurate and stable color.
}
\label{fig:main_comparison}
\end{figure*}

\clearpage
\fi

\begin{table*}[t]
\centering
\caption{
Quantitative comparisons on five scenes.
MAPE and LPIPS measure reconstruction quality, and time denotes inference time in milliseconds. We highlight the \rankonelegend{1st}, \ranktwolegend{2nd}, and \rankthreelegend{3rd} best in each column. Scene names are followed by the resolution of the rendered image. The results show that our method consistently achieves the best or second-best quality across all scenes, while also being the fastest among the compared methods.
}
\label{tab:main_quality}
\setlength{\tabcolsep}{3.2pt}
\renewcommand{\arraystretch}{1.08}
\small
\begin{tabular}{l|ccc|ccc|ccc|ccc|ccc}
\toprule
Method
& \multicolumn{3}{c|}{Bathroom (1024 $\times$ 1024)}
& \multicolumn{3}{c|}{Cornell-box (1024 $\times$ 1024)}
& \multicolumn{3}{c|}{Kitchen (1280 $\times$ 720)}
& \multicolumn{3}{c|}{Living-room (1280 $\times$ 720)}
& \multicolumn{3}{c}{Veach-ajar (1280 $\times$ 720)} \\
\cmidrule(lr){2-4}
\cmidrule(lr){5-7}
\cmidrule(lr){8-10}
\cmidrule(lr){11-13}
\cmidrule(lr){14-16}
& MAPE $\downarrow$ & LPIPS $\downarrow$ & Time $\downarrow$
& MAPE $\downarrow$ & LPIPS $\downarrow$ & Time $\downarrow$
& MAPE $\downarrow$ & LPIPS $\downarrow$ & Time $\downarrow$
& MAPE $\downarrow$ & LPIPS $\downarrow$ & Time $\downarrow$
& MAPE $\downarrow$ & LPIPS $\downarrow$ & Time $\downarrow$ \\
\midrule
Ours
& \rankonecell{0.091} & \rankonecell{0.084} & \rankonecell{14ms}
& \ranktwocell{0.059} & \rankonecell{0.023} & \rankonecell{11ms}
& \rankonecell{0.067} & \ranktwocell{0.079} & \rankonecell{20ms}
& \rankonecell{0.088} & \ranktwocell{0.100} & \rankonecell{11ms}
& \rankonecell{0.092} & \rankonecell{0.080} & \rankonecell{10ms} \\

NCR
& \ranktwocell{0.099} & \rankthreecell{0.095} & \rankthreecell{62ms}
& \rankonecell{0.058} & \rankthreecell{0.042} & 42ms
& \ranktwocell{0.075} & \rankthreecell{0.083} & \rankthreecell{73ms}
& 0.136 & 0.123 & 59ms
& \ranktwocell{0.112} & \ranktwocell{0.086} & \rankthreecell{39ms} \\

NR
& 0.131 & \ranktwocell{0.088} & \ranktwocell{28ms}
& \rankthreecell{0.074} & 0.048 & \ranktwocell{20ms}
& \rankthreecell{0.087} & 0.087 & \ranktwocell{44ms}
& \ranktwocell{0.124} & \rankthreecell{0.111} & \ranktwocell{23ms}
& 0.141 & \rankthreecell{0.104} & \ranktwocell{20ms} \\

OIDN
& \rankthreecell{0.125} & 0.106 & 71ms
& 0.076 & \ranktwocell{0.034} & \rankthreecell{38ms}
& 0.090 & \rankonecell{0.052} & 104ms
& \ranktwocell{0.124} & \rankonecell{0.083} & \rankthreecell{47ms}
& \rankthreecell{0.117} & 0.113 & 40ms \\

NR-RHS
& 0.378 & 0.699 & 229ms
& 0.279 & 0.632 & 125ms
& 0.347 & 0.691 & 475ms
& 0.561 & 0.754 & 173ms
& 0.514 & 0.762 & 145ms \\
\bottomrule
\end{tabular}
\vspace{-0.5em}
\end{table*}

\subsection{Representation Analysis}

To better understand what the learned representation stores, we further inspect OctaOctree itself rather than only the final rendered images.

\paragraph{Feature visualization.}
We visualize the learned features at different OctaOctree levels for a representative scene (See Figure~\ref{fig:feature_visualization}). The visualization reveals the complementary spatial-angular decomposition described in Section~3.4: coarse spatial levels capture broader reflected structures with stronger directional selectivity and low-frequency spatial patterns, while finer spatial levels focus on high-frequency local visibility changes, geometric detail, and short-range illumination variation. Together, these feature maps show that the hierarchy does not merely increase capacity, but organizes outgoing radiance according to the reuse pattern induced by glossy transport.

\ifshowmidfig
\begin{figure}[t]
\centering
\includegraphics[width=\columnwidth,height=0.7\textheight,keepaspectratio]{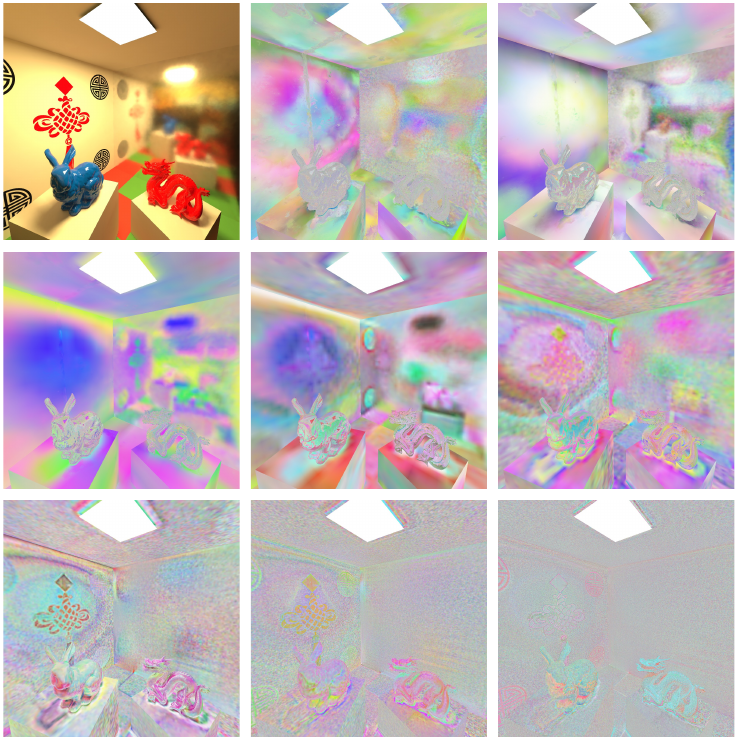}
\caption{
Visualization of learned features at different OctaOctree levels. Coarse spatial levels capture broader reflected structures with stronger directional selectivity, while finer spatial levels focus on local high-frequency changes, geometric detail, and short-range illumination variation.
}
\label{fig:feature_visualization}
\end{figure}
\fi

\paragraph{Structure statistics.}
We analyze the storage--quality trade-off of OctaOctree by varying the number of hierarchy levels on a representative scene. Table~\ref{tab:octa_stats} reports the maximum spatial and directional resolutions, total number of stored entries, feature storage, and the resulting reconstruction quality. Increasing the number of levels expands the complementary spatial-angular representation, providing finer spatial detail at high levels while preserving higher directional resolution at coarse levels. This generally improves image quality, but also increases storage cost. The results show that OctaOctree offers a controllable trade-off between memory usage and reconstruction fidelity, rather than relying on a fixed cache capacity.

\paragraph{Predicted disparity visualization.}
We visualize the predicted disparity for a representative scene (See Figure~\ref{fig:disparity_visualization}). The predicted disparity exhibits medium-frequency variation across the scene, which suggests that it adaptively distributes neighboring directional queries into different entries on the octahedral map. Meanwhile, smooth disparity distribution avoids abrupt changes in the reflected contents.

\ifshowmidfig
\begin{figure}[t]
\centering
\includegraphics[width=\columnwidth,height=0.7\textheight,keepaspectratio]{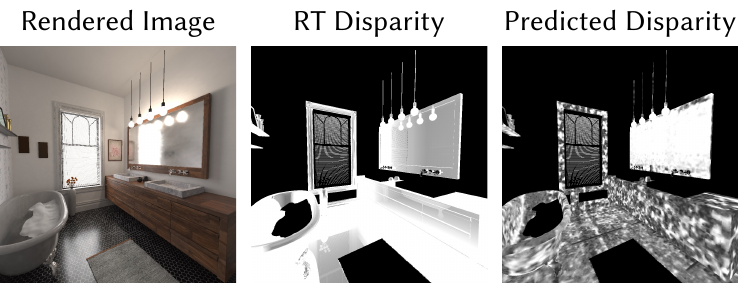}
\caption{
Visualization of ray-traced disparity and predicted disparity. The learned disparity is not a geometric hit distance, but a local reprojection field optimized for radiance reconstruction, which stabilizes neighboring octahedral lookups near reflection boundaries.
}
\label{fig:disparity_visualization}

\end{figure}
\fi

% \begin{table}[t]
% \centering
% \caption{
% OctaOctree statistics on a representative scene. We report the number of active octree nodes at each level, together with the total number of directional entries and the corresponding feature storage cost.
% }
% \label{tab:octa_stats}
% \begin{tabular}{cccc}
% \toprule
% Level & Spatial res. & Directional res. & Active nodes \\
% \midrule
% 1 & $2^2$ & $2^8+1$ & 56 \\
% 2 & $2^3$ & $2^7+1$ & 314 \\
% 3 & $2^4$ & $2^6+1$ & 1,585 \\
% 4 & $2^5$ & $2^5+1$ & 7,118 \\
% 5 & $2^6$ & $2^4+1$ & 30,690 \\
% 6 & $2^7$ & $2^3+1$ & 127,314 \\
% 7 & $2^8$ & $2^2+1$ & 583,036 \\
% 8 & $2^9$ & $2^1+1$ & 2,471,679 \\
% \midrule
% Total entries & \multicolumn{3}{c}{79,375,000} \\
% Feature storage & \multicolumn{3}{c}{1211 MB} \\
% \bottomrule
% \end{tabular}
% \vspace{-10pt}
% \end{table}

\begin{table}[t]
\centering
\caption{
OctaOctree statistics on a representative scene with different levels and resolutions. We report the maximum spatial/directional resolutions, number of total entries, feature storage, and image quality of varying resolution levels.
}
\label{tab:octa_stats}
\begin{tabular}{cccccc}
\toprule
Levels & Max Res. & Total entries & Storage & MAPE & LPIPS \\
\midrule
5 & $2^6/2^5$ & 734,275 & 11.2 MB & 0.178 & 0.106 \\
6 & $2^7/2^6$ & 3,526,245 & 53.8 MB & 0.149 & 0.096 \\
7 & $2^8/2^7$ & 16,957,768 & 258.8 MB & 0.125 & 0.092 \\
8 & $2^9/2^8$ & 79,375,000 & 1211 MB & 0.091 & 0.084 \\
\bottomrule
\end{tabular}
\vspace{-10pt}
\end{table}

\subsection{Ablation Study}

We perform ablations to evaluate the design choices specific to OctaOctree. In all cases, the full model uses mirror-reflection-centered lookup, predicted disparity for directional offset, octahedral barycentric interpolation, and spatial trilinear interpolation are applied.

\subsubsection{Directional Query Design}

The first group of ablations studies how the octahedral lookup direction is constructed (See Figure~\ref{fig:ablation_direction}).

`Outgoing dir.' replaces the mirror reflection direction $\omega_r$ with the viewing direction $\omega_o$ when accessing the octahedral map. `No direction offset' uses $\omega_r$ directly, without direction shifting. `Ray-traced depth offset' replaces the predicted disparity with the ray-traced depth measured along $\omega_r$. Results shows that using viewing direction directly or disabling direction shifting leads to blurry reflections and loss of detail, especially for glossy materials. Using ray-traced depth offset produces sharper reflections than the predicted disparity. Although this leads to slightly better MAPE, the result is still perceptually worse than the full model.

\ifshowmidfig
\begin{figure}[t]
\centering
\includegraphics[width=\columnwidth,height=0.7\textheight,keepaspectratio]{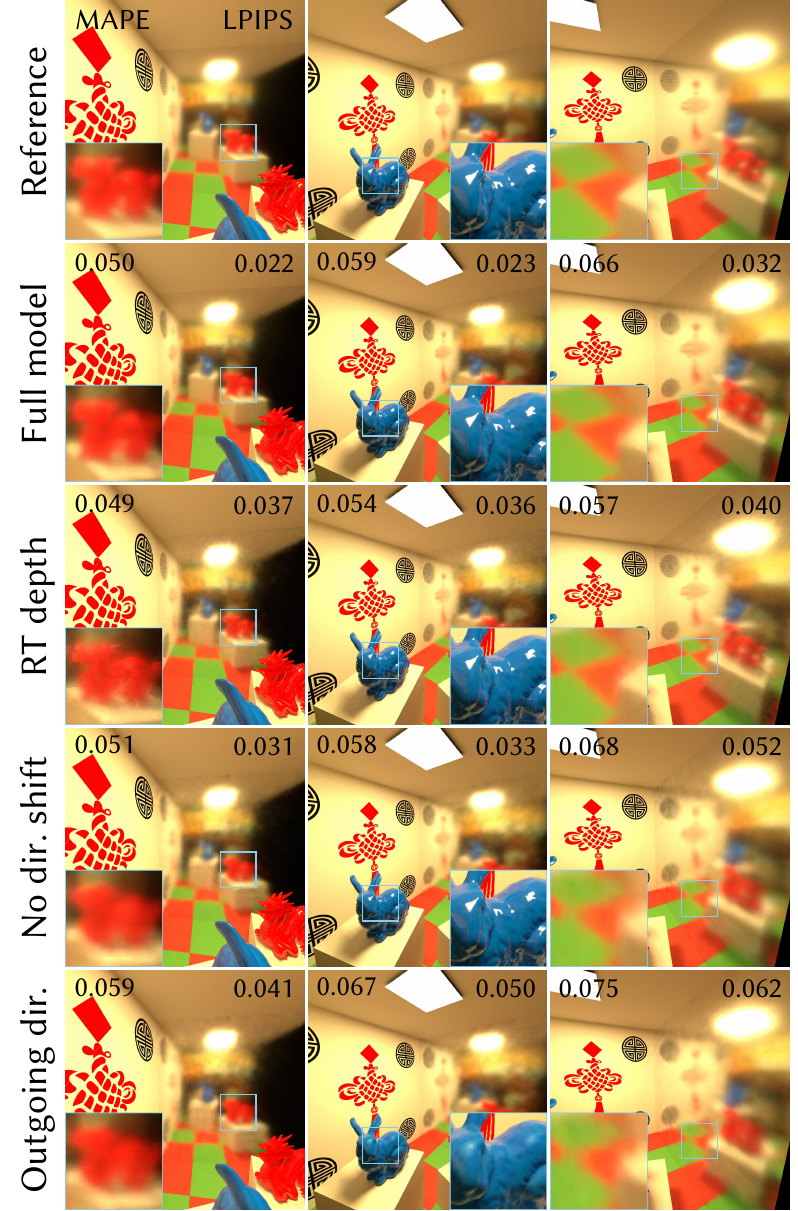}
\caption{
Ablation of directional lookup. We compare the full model with three variants: `Outgoing dir.' replaces the mirror reflection direction $\omega_r$ with the viewing direction $\omega_o$ when accessing the octahedral map. `No direction shift' uses $\omega_r$ directly, without direction shifting. `Ray-traced depth offset' replaces the predicted disparity with the ray-traced depth measured along $\omega_r$. Disabling direction shifting or using the viewing direction leads to blurry reflections and loss of detail. While using ray-traced depth offset produces better MAPE, it is perceptually worse than the full model due to abrupt changes in reflected contents. 
}
\label{fig:ablation_direction}
\end{figure}
\fi

% \begin{table}[t]
% \centering
% \caption{
% Ablation on directional query design. These variants test the importance of reflection-centered lookup and the predicted disparity offset.
% }
% \label{tab:ablation_query}
% \begin{tabular}{lcc}
% \toprule
% Method & MAPE $\downarrow$ & LPIPS $\downarrow$ \\
% \midrule
% Outgoing direction lookup & 0.067 & 0.051 \\
% No direction offset & 0.058 & 0.039 \\
% Ray-traced depth offset & \textbf{0.053} & 0.038 \\
% Full model & 0.058 & \textbf{0.026} \\
% \bottomrule
% \end{tabular}
% \end{table}

\subsubsection{Interpolation Strategy}

The second group studies the role of interpolation in the OctaOctree query (See Figure~\ref{fig:ablation_interpolation}).

% \begin{table}[t]
% \centering
% \caption{
% Ablation on OctaOctree interpolation. These variants test whether smooth interpolation is necessary in direction and space.
% }
% \label{tab:ablation_interp}
% \begin{tabular}{lcc}
% \toprule
% Method & MAPE $\downarrow$ & LPIPS $\downarrow$ \\
% \midrule
% No octahedral interpolation & 0.090 & 0.137 \\
% No octree interpolation & 0.143 & 0.230 \\
% Full model & \textbf{0.060} & \textbf{0.079} \\
% \bottomrule
% \end{tabular}
% \end{table}

Removing octahedral barycentric interpolation corresponds to nearest-vertex directional lookup on the octahedral map. Removing octree interpolation uses only the nearest spatial cell instead of trilinearly blending neighboring cells. Both variants reduce the smoothness of the queried representation and are expected to degrade reflected detail and highlight continuity.

\ifshowmidfig
\begin{figure}[t]
\centering
\includegraphics[width=\columnwidth,height=0.7\textheight,keepaspectratio]{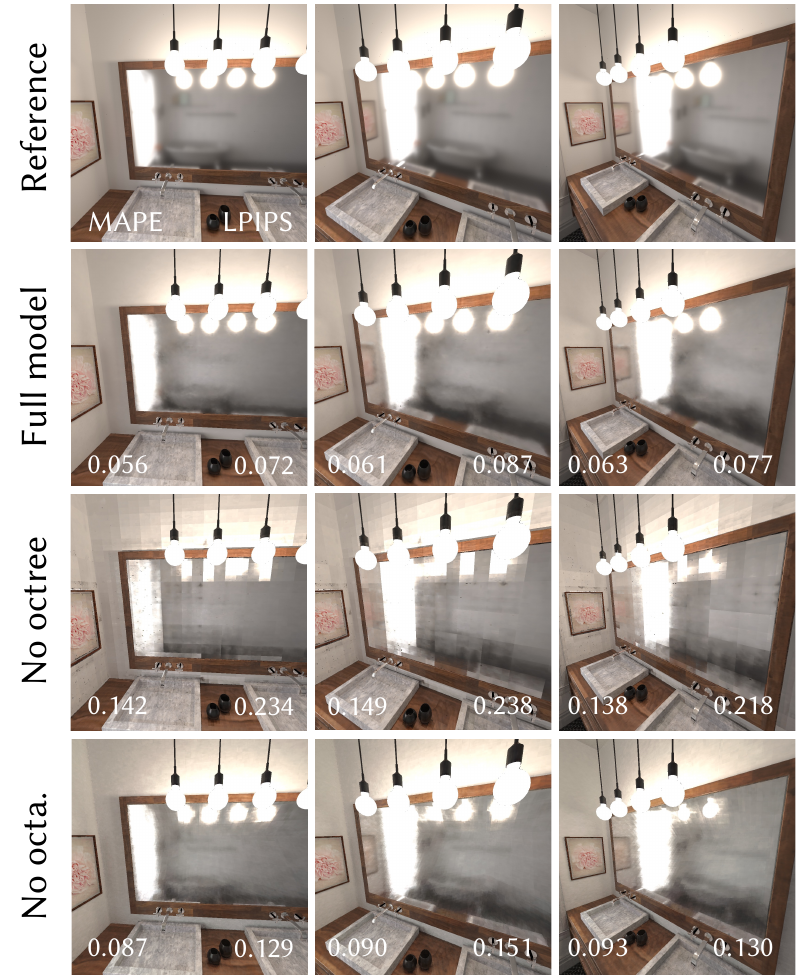}
\caption{
Ablation of spatial and angular interpolation. Without interpolation, the cache query directly uses the features stored at the nearest octree node and the closest direction bin, which leads to blocky artifacts and discontinuities in the predicted radiance. With smooth interpolation, the query blends features from multiple spatial nodes and direction bins, resulting in smoother and more accurate radiance estimates that better capture high-frequency variations.
}
\label{fig:ablation_interpolation}
\end{figure}
\fi

Overall, these ablations test the central claim of the paper: the quality of glossy neural radiosity depends not only on the decoder, but also on how outgoing radiance is organized, aligned, and interpolated inside the scene-space cache.

\section{Conclusion}

% We presented \emph{OctaOctree Neural Radiosity}, a scene-specific neural radiosity method built on an explicit spatial-angular representation of outgoing radiance. Our core idea is to organize a scene with a sparse octree in space and an octahedral map in direction, so that glossy and specular transport can be represented directly inside the cache rather than inferred only from positional features. By coupling this representation with disparity-aware directional lookup and a lightweight neural decoder, our method predicts outgoing radiance with a single cache query and decoder evaluation per visible shading point.

We presented \emph{OctaOctree Neural Radiosity}, a scene-specific neural radiosity method based on an explicit spatial-angular representation of outgoing radiance. By combining a sparse spatial octree with octahedral directional maps, our method stores glossy transport structure directly inside the representation instead of forcing compact neural networks to infer high-frequency directional variation implicitly. Together with disparity-aware directional lookup, OctaOctree enables efficient primary-hit radiance prediction with a single cache query and decoder evaluation.

% Our structured representation improves the reconstruction of glossy reflections, reflected detail, and other direction-dependent transport effects under limited sampling budgets. Compared with baseline methods, OctaOctree provides a more efficient way to encode high-frequency outgoing radiance without requiring multiple ray trace or network evaluation at inference time. Our analysis and ablations further indicate that the quality gain comes from the representation itself: complementary spatial-angular resolution, disparity-based alignment, and smooth interpolation all contribute to the final result.

The proposed structured representation improves glossy reflections and other direction-dependent transport effects while reducing the noise and overhead associated with repeated secondary-ray evaluation. Our analysis further suggests that the main benefit comes from the transport representation itself, including complementary spatial-angular resolution, reprojection-based alignment, and smooth interpolation.

% Our current formulation remains scene-specific, requiring a separate optimization for each target scene, and extremely specular reflections are still challenging. Even so, we believe OctaOctree demonstrates that structured spatial-angular caches are a promising direction for extending neural radiosity toward more accurate and efficient glossy global illumination.

Although our current formulation is evaluated only on static scenes, it is compatible with online optimization or dynamic neural radiosity frameworks~\cite{su2024dynamic} for handling scene changes. More broadly, the same spatial-angular organization may also be useful for other direction-dependent rendering problems, such as path guiding~\cite{dong2023neural} and control variates.

% \begin{acks}
% None.
% \end{acks}

\bibliographystyle{ACM-Reference-Format}
\bibliography{Sections/References}

\end{document}